\documentclass{aa}  
\usepackage{color}
\usepackage{graphicx}
\usepackage{lscape}
\usepackage{natbib}
\usepackage{natbib,twoopt}
\usepackage[varg]{txfonts}
\usepackage{url}
\usepackage{xspace}
\usepackage{amssymb}
\usepackage{stmaryrd}
\usepackage{siunitx}
\usepackage{textcomp}
\usepackage{pifont}

\usepackage[colorlinks = true,
linkcolor= blue,
citecolor = blue,
filecolor = blue,
urlcolor = blue,
]{hyperref}

\usepackage[usenames,dvipsnames]{xcolor}

\bibpunct{(}{)}{;}{a}{}{,}

\newcommand{\jnz}{\object{J0927$-$6335}\xspace}

\newcounter{Rco}

\newcommand{\logg}{\mbox{$\log g$}\xspace}

\newcommand{\Teff}{\mbox{$T_\mathrm{eff}$}\xspace}

\newcommand{\Msol}{$M_\odot$}

\defcitealias{El-Badry2023}{E23}

\begin{document}

\title{Ultraviolet spectroscopy of the supernova Ia hypervelocity runaway white dwarf J0927$-$6335}

\author{Klaus Werner\inst{1} \and Kareem El-Badry\inst{2} \and Boris T. G\"ansicke\inst{3} \and Ken J. Shen\inst{4}}

\institute{Institut f\"ur Astronomie und Astrophysik, Kepler Center for
  Astro and Particle Physics, Eberhard Karls Universit\"at, Sand~1, 72076
  T\"ubingen, Germany\\ \email{werner@astro.uni-tuebingen.de} 
\and
Department of Astronomy, California Institute of Technology, 1200 East California Boulevard, Pasadena, CA 91125, USA
\and
Department of Physics, University of Warwick, Coventry CV4 7AL, UK
\and 
Department of Astronomy and Theoretical Astrophysics Center, University of California, Berkeley, CA, USA
}

\date{Received 23 July 2024 / Accepted 12 August 2024}

\authorrunning{K. Werner et al.}
\titlerunning{UV spectroscopy of the SN Ia hypervelocity runaway WD J0927$-$6335}

\abstract{
The hot white dwarf (WD) \jnz (Gaia DR3 5250394728194220800, effective temperature \Teff = 60\,000 K, surface gravity \logg= 7) was detected as the fastest known Galactic hypervelocity star with a space velocity of $\approx2800$\,km\,s$^{-1}$ and an atmosphere dominated by carbon and oxygen. It is thought to be the surviving WD donor predicted by the ``dynamically driven double-degenerate double-detonation'' (D$^6$) type Ia supernova formation model. We analysed an ultraviolet spectrum of \jnz obtained recently with the \emph{Hubble Space Telescope} and found very high abundances of iron and nickel. This could originate in the pollution of the remnant by the SN Ia explosion but it is uncertain to what extent atomic diffusion altered the chemical composition of the accreted material.
}


\maketitle
%

\section{Introduction}
\label{sect:intro}

A population of hypervelocity white dwarfs (WDs) launched from thermonuclear supernovae (SNe) in close WD binaries was identified by \cite{Shen2018} through \emph{Gaia} astrometry. Additional such objects were found by \cite{El-Badry2023}. Unlike the cool and bloated objects discovered by \cite{Shen2018} , several of the sources identified by \cite{El-Badry2023} are hot, with \Teff $\gtrsim$ 60\,000\,K. One of them is \jnz (Gaia DR3 5250394728194220800), the fastest known star in the Galaxy with a space velocity of $\approx2800$\,km\,s$^{-1}$.  From an analysis of the optical spectrum, effective temperature and surface gravity of \Teff = $60\,000\pm5000$\,K and \logg = $7.0\pm0.5$ were found \citep{Werner2024}. Only lines of carbon and oxygen were identified (ionisation stages \ion{C}{iii-iv} and \ion{O}{iii-iv}) and it was inferred that the atmosphere is composed of carbon and oxygen in equal amounts (by mass). An upper limit of the helium abundance was derived (He $<$ 0.05). It was concluded that \jnz is a runaway WD that was eroded by the accretor both prior to and during its explosion. \jnz and other similar hypervelocity WDs were designated D$^6$ stars after their so-called ``dynamically driven double-degenerate double-detonation'' formation model \citep{Shen2018}. It can be expected that the surviving WD is polluted by the explosion debris from the companion which are predicted to consist mostly of iron-group elements \citep[e.g.,][]{Keegans2023}. Because of the high effective temperature, however, heavy metals can only be detected in ultraviolet (UV) spectra. 

We therefore observed \jnz with the \emph{Hubble Space Telescope} (HST) and conducted a model-atmosphere analysis in order to test the D$^6$ formation scenario for hypervelocity WDs. Abundance measurements may constrain the yields of the SNe from which these stars were born and the masses of the exploded WDs. That could help to shed light on the debated physics and origin of type Ia SNe \citep[e.g.,][]{liu2023,Shen2024}.

\section{Observation}
\label{sect:observation}

\begin{figure*}
  \centering  \includegraphics[width=1.00\textwidth]{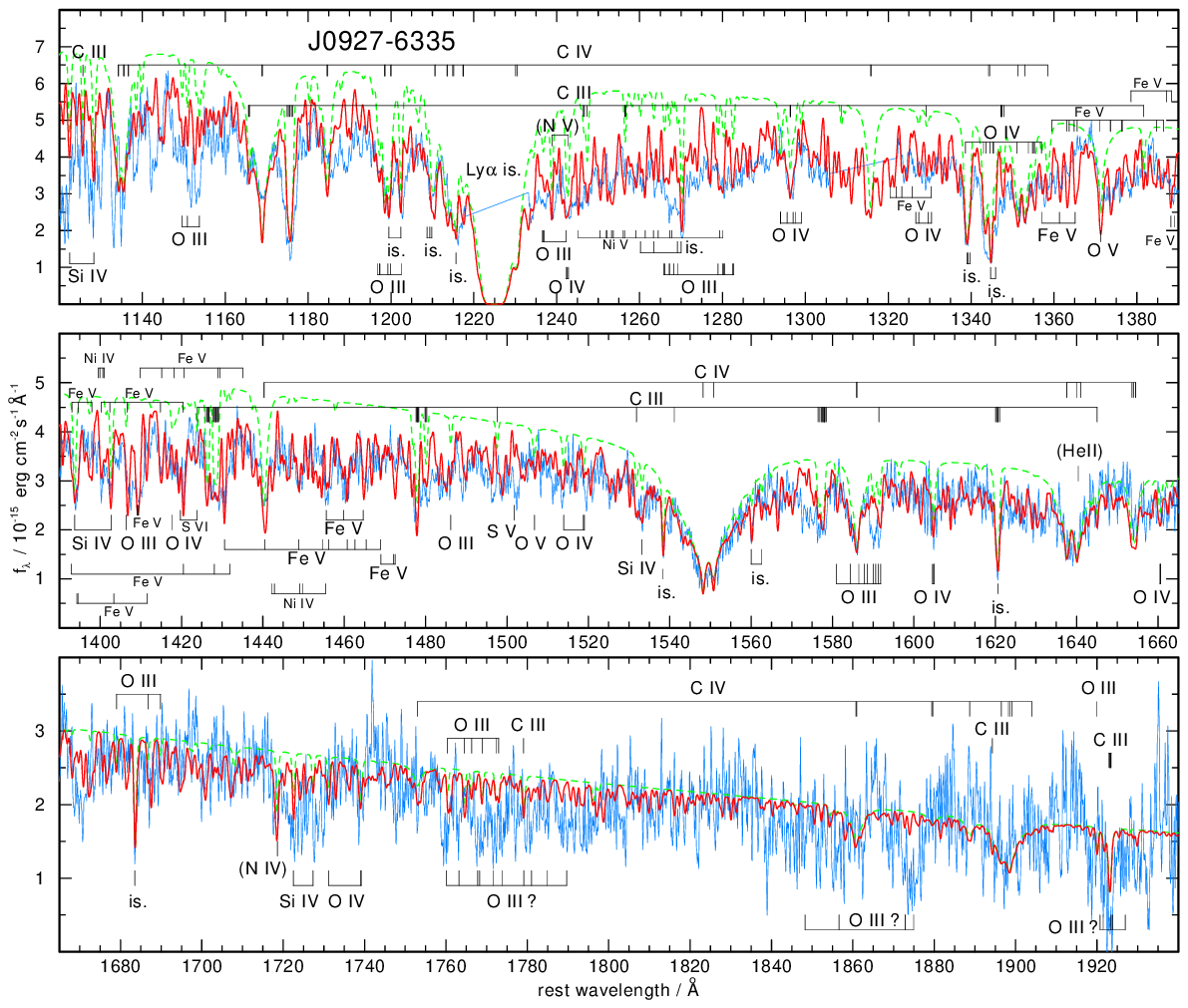}
 \caption{HST spectrum of \jnz (blue graph) shifted to rest wavelength and compared to our final model (red graph) with \Teff = 60\,000\,K, \logg = 7.0, C = 0.42, N = 0.0007, O = 0.42, Si = 0.02, Fe = 0.10, and Ni = 0.03 (mass fractions). The spectrum is dominated by strong \ion{C}{iii}/\ion{C}{iv} lines and weaker \ion{O}{iii}/\ion{O}{iv} lines as well as \ion{O}{v}~1371\,\AA. The absorption feature at 1640\,\AA\ is not caused by \ion{He}{ii} but by \ion{C}{iv}. The continuum flux distribution is strongly affected by a curtain of numerous overlapping Fe and Ni lines, as can be judged from the green dashed graph which is the same model but without Fe and Ni line opacities. Only a few of the strongest multiplets of \ion{Fe}{v}, \ion{Ni}{iv}, and \ion{Ni}{v} can be identified individually. Labels ``is.'' denote interstellar absorption lines. They are also included in the model spectrum. }
\label{fig:J0927-6335}
\end{figure*}

\jnz was observed with the Cosmic Origins Spectrograph (COS) aboard HST on April 9, 2024 (proposal ID 17441, PI El-Badry). Grating G140L with a spectral resolution of $\Delta\lambda = 0.6$\,\AA\ was used with central wavelength set to 1105\,\AA. The spectrum covers the wavelength range $\approx 1120-2250$\,\AA, however, the strongly decreasing signal-to-noise (S/N) ratio towards the red end limits the useful range to about 1940\,\AA. The spectrum was smoothed with a 0.2\,\AA\ wide boxcar to increase the S/N ratio without significant loss of spectral resolution. We cut out strong airglow lines from \ion{H}{i} Ly$\alpha$ at 1216\,\AA\ and \ion{O}{i} near 1305\,\AA. The spectrum is displayed in Fig.\,\ref{fig:J0927-6335}. It was shifted to rest wavelengths accounting for the radial velocity of the star ($v_{\rm rad}=-2285$\,km s$^{-1}$) as determined by \cite{El-Badry2023} based on higher-resolution optical spectroscopy. We also tried measuring the star's radial velocity directly from the UV spectrum and found  $v_{\rm rad}=-2265 \pm 20$\,km s$^{-1}$, fully consistent with the value inferred from the optical spectra.

\begin{figure}
  \centering  \includegraphics[width=1.00\columnwidth]{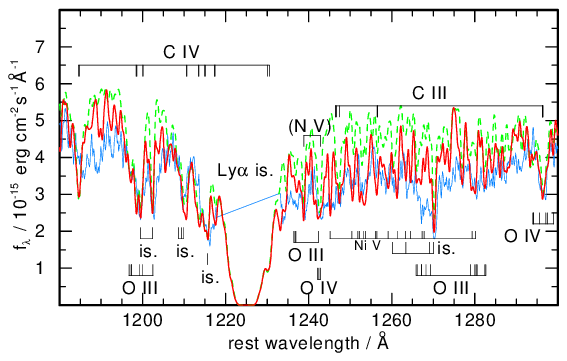}
 \caption{Similar to Fig.\,\ref{fig:J0927-6335}. The final model (red graph) with Fe = 0.10 and Ni = 0.03 is compared to a model with Fe = 0.30 and Ni = 0 (green dashed graph).}
\label{fig:J0927-6335_detail}
\end{figure}

\begin{table*}[t]
\begin{center}
\caption{Atmospheric element abundances of \jnz and statistics of NLTE model atoms.
\tablefootmark{a} 
}
\label{tab:results}
\begin{tabular}{ccccccccccc}
\hline 
\hline 
\noalign{\smallskip}
Element & Mass Fraction    & [X]     & I     & II     & III      & IV      & V        & VI    & VII    & VIII \\
\hline
\noalign{\smallskip}
He      & $<0.05$          & $<-0.7$ & 29, 60& 15,105 & 1, 0  \\   
C       & $0.42 \pm 0.25$  & 2.3     &       & 1, 0   & 133, 745 & 54, 279 & 1, 0    \\ 
N       & $< 0.0007$       & $<0.0$  &       &        & 1, 0     & 76, 405 & 54, 297  & 1, 0  \\ 
O       & $0.42 \pm 0.25$  & 1.9     &       &        & 47, 171  & 83, 637 & 105, 671 & 9, 15  & 1, 0 \\ 
Si      & $0.02 \pm 0.01$  & 1.5     &       &        & 17, 28   & 30, 102 & 25, 59   & 1, 0   \\ 
S       & $< 0.005$        & $<1.2$  &       &        &          & 17, 32  & 39, 107  & 25, 48 & 1, 0\\ 
Cr      & $<0.01$          & $<2.8$  &       &        &          & 7, 23   & 7, 24    & 7, 23  & 7, 24 & 1, 0 \\ 
Mn      & $<0.003$         & $<2.4$  &       &        &          & 7, 25   & 7, 25    & 7, 24  & 7, 23 & 1, 0 \\ 
Fe      & $0.10 \pm 0.05$  & 1.9     &       &        &          & 7, 25   & 7, 25    & 7, 25  & 7, 24 & 1, 0 \\ 
Ni      & $0.03 \pm 0.02$  & 2.6     &       &        &          & 7, 25   & 7, 27    & 7, 27  & 7, 25 & 1, 0 \\
\hline
\end{tabular} 
\tablefoot{  
\tablefoottext{a}{[X] denotes the logarithm of the mass fractions relative to the solar values from \cite{2009ARA&A..47..481A}. Columns I--VIII list the number of NLTE levels and lines (first and second number of each table entry, respectively) of the respective ionization stages of the model atoms. For Cr, Mn, Fe, and Ni, these are superlevels and superlines; see text.} } 
\end{center}
\end{table*}

\begin{table}[t]
\begin{center}
\caption{Approximate spectral regions where Fe and Ni lines are predominantly located.
}
\label{tab:lines}
\begin{tabular}{ccccccccccc}
\hline 
\hline 
\noalign{\smallskip}
\ion{Fe}{iv}& \ion{Fe}{v}& \ion{Fe}{vi}&\\
1400--1900\,\AA& 1100--1600\,\AA& 1230--1340\,\AA\\
\noalign{\smallskip}
 \ion{Ni}{iv}& \ion{Ni}{v}& \ion{Ni}{vi}&\\
1300--1600\,\AA& 1100--1600\,\AA& 1100--1330\,\AA\\
\hline
\end{tabular} 
\end{center}
\end{table}

\begin{table*}[t]
\small
\begin{center}
\caption{Element abundances in \jnz compared to other compact objects.
\tablefootmark{a} 
}
\label{tab:abu}
\begin{tabular}{cccccccccccc}
\hline 
\hline 
\noalign{\smallskip}
Star & Type   &   \Teff/\logg & H  &  He             &  C              & O               & Si             & S              & Fe               & Ni               \\
\hline
\noalign{\smallskip}
\jnz          &D6 &60\,000/7.0& -- & $<0.05$         & 0.42            & 0.42            & 0.02           & $<0.005$       & 0.10             & 0.03             \\
G191-B2B      &DA &60\,000/7.6&0.99&$<2\times10^{-5}$&$7\times10^{-6}$ &$2\times10^{-5}$ &$5\times10^{-5}$&$6\times10^{-6}$&$5\times10^{-4}$  &$3.5\times10^{-5}$\\
EC11481$-$2302&sdO&55\,000/5.8&0.87&0.009            &$<8\times10^{-9}$&$<1\times10^{-6}$& --             & --             & 0.045            & 0.025            \\
HS0713+3958   &DO &65\,000/7.5& -- & 0.99            &$7\times10^{-6}$ &$<3\times10^{-6}$&$5\times10^{-6}$&$1\times10^{-5}$&$1.3\times10^{-4}$&$5\times10^{-4}$  \\
\hline
\end{tabular} 
\tablefoot{  
\tablefoottext{a}{Data for G191-B2B, EC11481$-$2302, HS0713+3958 from \cite{Rauch2013}, \cite{Landstorfer2024}, \cite{Werner2018}, respectively. Abundances in mass fractions.
} } 
\end{center}
\end{table*}

\section{Spectral analysis}
\label{sect:analysis}

We used the T\"ubingen Model-Atmosphere Package (TMAP) to compute non-local thermodynamic equilibrium (NLTE), plane-parallel, line-blanketed atmosphere models in radiative and hydrostatic equilibrium \citep{2003ASPC..288...31W}. The chemical constituents of the initial model which was constructed to match the optical spectrum of \jnz are helium, carbon, and oxygen. See \cite{Werner2024} for details. For the present analysis of the UV spectrum, a new set of models was computed, including additionally iron and nickel. In order to account for the large number of iron and nickel lines, we employed a statistical approach (using so-called superlevels and superlines) which enables the construction of models under NLTE conditions \citep{Anderson1989,DreizlerWerner1993,Rauch2003}. In essence, we included lines from Kurucz's lists \citep{kurucz2009} for \ion{Fe}{iv-vii} and \ion{Ni}{iv-vii}. These so-called LIN lists comprise in total 7,561,008 and 16,882,369 lines with laboratory-observed and computed wavelengths from iron and nickel, respectively, in the considered ionisation stages. The synthetic spectra which are compared to the observation were computed with the so-called POS lists, which are a subset of the LIN lists containing laboratory-observed lines only. Subsequent NLTE line formation iterations (i.e., keeping fixed the atmospheric structure) were carried out for nitrogen, silicon, and sulfur as well as for chromium and manganese. The latter two are numerically treated like iron and nickel. Table~\ref{tab:results} summarizes the used model atoms. In order to reduce computational time, model atoms smaller than listed in the table were used for some ions when calculating the atmospheric structure. These are \ion{C}{iii} (44 NLTE levels, 175 lines), \ion{O}{iii} (1, 0), \ion{O}{iv} (38, 173), and \ion{O}{v} (34, 114).

We started the analysis of the HST spectrum by extending the synthetic spectrum from the final model of our optical analysis to the UV. That helped us to identify the strongest observed lines which stem from carbon and oxygen. Like in the optical spectrum, these are lines from \ion{C}{iii} and \ion{C}{iv} -- with the 1548/1551\,\AA\ resonance line as the most prominent in the entire spectrum -- as well as \ion{O}{iii} and \ion{O}{iv}. In contrast to the optical spectrum, we can also detect a \ion{O}{v} line, namely that at 1371\,\AA\ (but it is blended with a strong \ion{Fe}{v} line). The relative strength of the oxygen lines from all three ionization stages is well reproduced by the model, indicating that the effective temperature of the model (\Teff = 60\,000\,K) correctly predicts the ionization balance. The same holds for the relative strengths of the carbon lines from both ionization stages. Also, the absolute strengths of the carbon and oxygen lines are matched well, indicating that carbon and oxygen abundances are similar, as found from the optical spectrum. The broad wings of \ion{C}{iv} 1548/1551\,\AA\ are well reproduced by the model, showing that surface gravity found from the optical line analysis (\logg=7) is correct. For the following analysis steps, we therefore kept fixed \Teff = 60\,000\,K, \logg = 7, and C/O = 1.

We reddened our models by using the color excess $E(B-V) = 0.15$, found by a fit to the spectral energy distribution of \jnz \citep{Werner2024}, and assuming a \cite{Fitzpatrick1999} reddening law with $R_{\rm v}=3.1$. The same value for the reddening was adopted by \cite{El-Badry2023}, considering the predicted values of 0.13 and 0.2 for infinity from the \cite{Lallement+2022} 3D dustmap and the \cite{Schlegel1998} dustmap, respectively. Using the relation  $\log(N_{\rm H}/E(B-V))=21.58$ \citep{GroenewegenLamers1989} one obtains the interstellar neutral hydrogen column density $n_{\rm H}= 5.7\times10^{20}$\,cm$^{-2}$. With this value, our models are attenuated by the resulting broad interstellar Lyman\,$\alpha$ line. Note that it does not appear centered at its rest wavelength at 1216\,\AA\ in Fig.\,\ref{fig:J0927-6335}, but redward at about 1225\,\AA, because we shifted the stellar spectrum to rest wavelengths. We identified a few other commonly occurring interstellar lines which we also modeled in the synthetic spectrum by freely customizing the ion column densities. They are indicated in Fig.\,\ref{fig:J0927-6335}. We found no clear hints for the presence of interstellar resonance line doublets of \ion{C}{iv}, \ion{N}{v}, and \ion{Si}{iv}, however, strong photospheric metal line blanketing might preclude their detection.

It is worth to remark that the absorption feature at 1640\,\AA\ is not caused by photospheric helium but by carbon. We see a blend of some components of a \ion{C}{iv} transition between atomic levels with principal quantum numbers $n = 4 \rightarrow 6$ of this lithium-like ion, instead of the $n=2\rightarrow3$ in the isoelectronic \ion{He}{ii} ion. The situation is similar to that in the optical region, where \ion{C}{iv} $n=6\rightarrow8$ transitions mimic a \ion{He}{ii} $n=3\rightarrow4$ line at 4686\,\AA. The HST spectrum therefore confirms the helium-deficiency of \jnz (see below).

Aside from the good representation of the carbon and oxygen lines, however, the original C/O model of the optical analysis of \cite{Werner2024} did not match the UV continuum flux distribution. It significantly overpredicts the flux of the COS spectrum in the short-wavelength region up to about 1640\,\AA. This cannot be solved by stronger reddening. Increasing $E(B-V)$ from 0.15 to 0.2 would help only at the shortest wavelengths of the spectrum. In addition, the observed overall flux distribution shows structures that indicate additional line opacity sources. This situation is reminiscent of a similar situation in the UV spectra of some hot subdwarf stars, where heavy line blanketing by overabundant iron-group elements was found, e.g., in the hydrogen-rich sdO star EC\,11481-2303 \citep[\Teff = 55\,000, \logg = 5.8,][]{RingatRauch2012,Landstorfer2024}. The heavy-element excess in hot subdwarfs is the consequence of atomic diffusion driven by selective radiative levitation.

Iron and nickel have a large number of lines in the UV spectra of hot stars. Given the moderate resolution of the HST/COS spectrum of \jnz, only the strongest ones can be identified individually. They stem from \ion{Fe}{v} and \ion{Ni}{iv-v} and some of them are marked in Fig.\,\ref{fig:J0927-6335}. But it is the majority of lines of iron and nickel in ionisation stages IV--VI with weaker opacity which effectively suppress the continuum flux, especially when the element abundances are high. A comparison of our final model with a synthetic spectrum from which we removed all iron and nickel lines shows that line blanketing in the short-wavelength region up to about 1600\,\AA\ is so severe, that there are only a few narrow windows with no or very weak lines, where we see the true continuum. One example can be seen just blueward of the \ion{O}{v}~1371\,\AA\ line in the region 1365--1370\,\AA\ (top panel of Fig.\,\ref{fig:J0927-6335}), which at first glance looks like an emission feature. The flux on the red side of the \ion{O}{v} line core is strongly absorbed by a forest of many strong \ion{Fe}{v} lines.

By switching off lines from single ions in the synthetic spectrum, we found that different ions of iron and nickel are blocking the flux in different, however overlapping, wavelengths regions. Generally, with increasing ionisation stage the lines appear at shorter wavelengths and the contribution of nickel to the opacity is stronger than that of iron. In Tab.\,\ref{tab:lines} we indicate in which wavelength region we find the strongest contribution of relevant iron and nickel ions. 

We calculated about a dozen models with various iron and nickel abundances and picked by eye the best fitting one and estimated the errors by excluding models that show significantly worse fits. The final model has Fe = $0.10\pm0.05$ and Ni = $0.03\pm0.02$. The error limits are quite large, because it is difficult to disentangle the contributions of iron and nickel to the spectral energy distribution. To some extent, a higher abundance of one element can be compensated by a lower abundance of the other. However, Fig.\,\ref{fig:J0927-6335_detail} shows a spectral region where the complete lack of nickel cannot be compensated even if iron is increased to Fe = 0.30, i.e., a factor of three higher than in our best-fit model.

The only other species besides C, O, Fe, and Ni which we identified are silicon and perhaps sulfur. It is possible to recognize the resonance doublet of \ion{Si}{iv} at 1394/1403\,\AA, although it is located in a wavelength region where the strongest \ion{Fe}{v} multiplets are found. Model fits with various silicon abundances arrive at  Si = $0.02 \pm 0.01$. A few other \ion{Si}{iv} lines are visible in our model and they are labelled in Fig.\,\ref{fig:J0927-6335}. However, they are weaker than the resonance doublet and cannot be identified in the observation. As to sulfur, we can constrain the abundance using the strongest \ion{S}{v} line which is located at 1501.76\,\AA. Unfortunately, it is blended by an unidentified line located redward by about 0.5\,\AA. By varying the sulfur abundance we find an good fit at S = $0.005$, however, we regard this value just as an upper limit. A \ion{S}{v} multiplet is located at 1128--1134\,\AA, but its strongest components are blended by dominating \ion{Si}{iv} lines. A doublet of \ion{S}{vi} is at 1420/1424\,\AA\ which, however, is also blended by other lines. At least we can use it to confirm that S $<0.005$. 

We put upper abundance limits to a few other species. For nitrogen we used the absence of the \ion{N}{v}~1239/43 resonance doublet as well as the \ion{N}{iv}~1719\,\AA. We find N $< 7 \times 10^{-4}$, which is the solar mass fraction \citep{2009ARA&A..47..481A}. We also confirmed the upper limit of the helium abundance (He $< 0.05$) found previously from the optical spectrum. At higher abundances, a \ion{He}{ii} line core at 1640\,\AA\ appears in the models in contrast to the observation. As to the iron-group elements chromium and manganese, upper abundance limits were determined, too. For both species, a small number of the strongest predicted lines (\ion{Cr}{v}, \ion{Cr}{vi}, \ion{Mn}{v}, \ion{Mn}{vi}) were used for this purpose. We find Cr $< 0.01$ and Mn $< 0.003$. Table\,\ref{tab:results} summarizes our abundance analysis.

There are some absorption features in the HST spectrum which are not reproduced by our model and which we are unable to identify, see for instance at 1573\,\AA. At longer wavelengths (bottom panel of Fig.\,\ref{fig:J0927-6335}) we see more unidentified absorptions, e.g., at 1670\,\AA. Maybe they are spurious features because of the low S/N but on the other hand, a similar strong feature at 1921\,\AA\ is a \ion{C}{iii} line closely matched by the model. Some unidentified lines could stem from \ion{O}{iii}. They are located near 1760--1790\,\AA, 1850--1875\,\AA, and 1920--1925\,\AA, and we marked them with the label ``\ion{O}{iii}?'' in Fig.\,\ref{fig:J0927-6335}. These lines appear in the atomic spectra database\footnote{\url{https://www.nist.gov/pml/atomic-spectra-database}} of the National Institute for Standards and Technologies \citep[NIST,][]{Kramida2020} and were first reported as strong lines in laboratory spectra by \cite{Bockasten1964}. They are particular lines because for the upper levels involved, intermediate coupling is significant. This is probably one reason why no oscillator strengths were computed and therefore we cannot model these lines.

\section{Summary and discussion}
\label{sect:summary}

We found that the carbon and oxygen dominated atmosphere of \jnz (C/O=1) is strongly enriched in iron and nickel, namely, Fe = $0.10\pm 0.05$ and Ni = $0.03\pm0.02$. We can interpret this composition as a mixture of C and O of the escaping CO WD and deposited SN ejecta from the exploded WD. The absolute abundances of Fe and Ni depend on the mixing of the contaminants in the C/O envelope. But we can compare the observed Ni/Fe ratio (0.067--1.0) to model predictions for sub-Chandraskhar explosions, which depend on the WD mass $M$ and metallicity $Z$. For solar-like metallicities and depending on the WD mass, the Ni/Fe ratio is between about 0.003--0.06 \citep{Shen2018b}. The predictions are smaller than the value we found for \jnz albeit the lower limit of our result is marginally consistent with the prediction for a $Z=2$ and a M = 1.1\,\Msol\ WD. 

Iron and nickel abundances determined from X-ray spectroscopy of two evolved Ia supernova remnants (SNRs) resulted in Ni/Fe ratios which are compatible with the mentioned predictions of sub-Chandrasekhar explosions \citep{Yamaguchi2015,Shen2018b}. For the Kepler and Tycho SNRs, Ni/Fe = 0.03--0.07 and 0.02--0.04, respectively, were found \citep{Park2013,Yamaguchi2014,Yamaguchi2015}, pointing at $Z\approx 2$ and M $\approx1$\Msol\ WD progenitors. A significantly larger ratio (Ni/Fe = 0.11--0.24) which is closer to the value we measured for \jnz, was found for the SNR 3C\,397 \citep{Yamaguchi2015}. Such a high value would require unrealistically high initial metallicity for a sub-Chandrasekhar SN and thus was claimed as evidence for a Chandrasekhar-mass explosion, i.e., a single-degenerate (SD) origin of 3C\,397. But \cite{Shen2018b} argue that the SD scenario also has problems to explain the high Ni/Fe ratio, concluding that this object continues to present a nucleosynthetic puzzle for any SN type Ia scenario. In this sense, the WD of our study represents another such puzzle piece.

We briefly mentioned the sdO star EC11481$-$2302, in which the strong enrichment in Fe and Ni is caused by radiative levitation. So the question arises whether this process is relevant for \jnz, too. In Tab.\,\ref{tab:abu} we compare the element abundances of these two stars as well as a DA and a DO WD with similar temperature and gravity like \jnz. Only the sdO star shows an extreme Fe and Ni enrichment like \jnz, but its surface gravity is much lower. Silicon is very much lower in the DA and in the DO. That could mean that gravitational settling and radiative levitation is not at work in \jnz. But in contrast to H- and He-dominated objects, we do not know whether and how atomic diffusion acts on Si, Fe, and Ni in a CO-dominated atmosphere. Investigations of SN~Iax remnant models by \cite{ZhangFuller2019} predict that the atmospheres are depleted of C and O and dominated by Fe and Ni because of radiative levitation (their fig.\,9), however, the authors emphasized that this extreme effect is probably unrealistic due to their inaccurate numerical treatment of diffusion. In addition, they disregarded stellar winds, which hamper the effects of levitation and settling \citep{Unglaub2000}. So we are left with the conclusion the high abundances of Fe and Ni in \jnz could signal the pollution of the remnant by the SN Ia explosion but we are uncertain to what extent diffusion altered the chemical composition of the accreted material.

\begin{acknowledgements} 
This research is based on observations made with the NASA/ESA Hubble Space Telescope obtained from the Space Telescope Science Institute, which is operated by the Association of Universities for Research in Astronomy, Inc., under NASA contract NAS 5–26555. These observations are associated with program ID 17441. The TMAD tool (\url{http://astro.uni-tuebingen.de/~TMAD}) used for this paper was constructed as part of the activities of the German Astrophysical Virtual Observatory. This research has made use of NASA's Astrophysics Data System and the SIMBAD database, operated at CDS, Strasbourg, France. This research has made use of the VizieR catalogue access tool, CDS, Strasbourg, France.  K.E. was supported in part by HST-GO-17441.001-A. This project has received funding from the European Research Council (ERC) under the European Union’s Horizon 2020 research and innovation programme (Grant agreement No. 101020057).
\end{acknowledgements}

\bibliographystyle{aa}
\bibliography{aa}

\end{document}